\documentclass[12pt]{article}
\usepackage{epic}
\usepackage{rotating}
\usepackage{subfigure}
\usepackage{psfig}
\usepackage{epsfig}
\usepackage{wrapfig}
\newcommand{\ra}{\rangle}
\newcommand{\la}{\langle}
\newcommand{\Tm}{\T^{\mathrm (m)}}
\newcommand{\T}{T_{\mathrm eff}}
\title{Approximate real time visualization of a Rabi transition by
means of continuous fuzzy measurement}
\author{
\large J\"urgen Audretsch\thanks{E-mail: 
Juergen.Audretsch@uni-konstanz.de} \and  
Thomas.Konrad\thanks{E-mail: Thomas.Konrad.@uni-konstanz.de} \and
Michael Mensky\thanks{Permanent address:
P.N.Lebedev Physical Institute, 117924 Moscow, Russia.\newline
E-mail: mensky@sci.lebedev.ru}
\\
\normalsize \it Fakult\"at f\"ur Physik der Universit\"at Konstanz\\
\normalsize \it Postfach 5560 M 674, D-78434 Konstanz, Germany}
\date{07.07.99}
\begin{document}
\maketitle
\begin{abstract}
Continuous weak or fuzzy measurement of the Rabi oscillation of a two
level atom 
subjected to a $\pi-$pulse of a resonant light field is simulated
numerically. We thereby address the question whether it is possible to
measure characteristic features of the motion of the state of a single quantum
system in real time. We compare two schemes of continuous measurement:
continuous measurement with constant fuzziness and with 
fuzziness changing in the course of the measurement. Because the
sensitivity of the Rabi atom to the influence of the measurement
depends on the state of the atom, it is possible to optimize the
continuous fuzzy measurement by varying its fuzziness.  
\end{abstract}

\section{Introduction}
We consider a single two-level atom with energy eigenstates $|E_1\ra$
and $|E_2\ra$ under the influence of  a time dependent potential
$V(t)$. Assuming $V(t)$ is not known  and the evolution of the atom in the
potential can be observed only once, which scheme of measurement conveys the most
information about the otherwise undisturbed motion of the state $|\psi(t)\ra$ of
the atom in the potential?
We will restrict to a reduced information and ask how to optimally record
with minimal disturbance the evolution in time of the squared modulus of one of the components of the state of the atom, say
$|c_2(t)|^2$, where $|\psi(t)\ra
= c_1(t)|E_1\ra + c_2(t)|E_2\ra$ is the normalized state vector.
What types of measurement are to our disposition?\\
In order to detect the evolution of $|c_2(t)|^2$ by means of
projection measurements of energy,  
the system has to be prepared new before each measurement in the same
initial state with the same potential $V(t)$.
If a  sequence of consecutive projection measurements is
applied because the system cannot be prepared new, the
dynamics of the atom are in general strongly altered \cite{Milburn}. In the
continuum limit of an infinite sequence of projection measurements the
evolution can even be halted (Quantum Zeno effect) \cite{Zeno}. Therefore
 the usual projection measurements are not an appropriate choice to detect
characteristic features of the evolution of the atomic state in real-time
(i.e. without resetting the system).
 
In fact for this purpose two 
requirements have to be fulfilled simultaneously: the influence of the measurement
on the dynamics of the system must not be too strong and the
measurement readout has to be
accurate enough to indicate the evolution of the state. 
The difficulty lies in the unavoidable competition of these properties:
the better the dynamics are conserved (i.e. the smaller the influence of the
measurement), the less reliable is the readout and vice versa. 

We will show below that an approach to a real-time visualization can
profitably be
based on 
unsharp measurements which for example can be described in the POVM formalism
\cite{Ali74}. Unsharp measurements are there associated with observables that can not be
represented as projection valued measures but only as positive
operator valued measures. These measurements have the advantage that
their influence is less strong than the influence of projection
measurements, but at the same time they have a lower resolution.
In order to obtain information about $|c_2(t)|^2$ in real-time, if only one realization of the evolution is
available, a whole sequence of unsharp measurements is
required. For calculational convenience we 
consider such sequences in the limit of continuous 
measurements.  

A continuous measurement of energy lasts over a certain time period
($0\le t\le T$) and produces a
readout denoted by $[E] =\{(t,E(t))|0\le t\le T\}$ which assigns to each time $t$ during this period a measured
value $E(t)$. It serves to detect  the evolution of a quantum mechanical
system.
Continuous measurements have been investigated in several
contexts \cite{ContinMeas}.  
Since we use unsharp measurements the readout possesses in
general a low accuracy and shows quantum fluctuations. Therefore these
measurements are called 
continuous {\it fuzzy} measurements. 
We will base our considerations below on a phenomenological model,
which also makes plausible that a correlation between  $E(t)$ and
$|c_2(t)|^2$ is to be expected. For a realization scheme see
\cite{AudM98}.

With continuous fuzzy measurements we have found a possible candidate
for the measurement scheme with the desired properties. But note, that
according to the nature of quantum mechanics, there is no scheme that precisely
records  the evolution of $|c_2(t)|^2$ in real-time without influencing
it as well.  All we can do  is to find a mechanism, that allows a ``best bet''
on the behaviour of $|c_2(t)|^2$, if only one ``run'' is available.  

The relation between the modification of the motion of the state and the
reliability of a readout has been  
investigated in \cite{AudM98,AudM97, AuMNam97}. There it was shown for
measurements with constant fuzziness, that a
visualization can be achieved up to a certain degree if the value of
fuzziness is properly chosen.
In the present work we want to extend the investigation to the case
where the parameter which determines the fuzziness of the measurement
is varied as function of time or is made dependent on the readout
$E(t)$ of the
measurement. Our intention is to show that in this way it is possible to improve
the efficiency
of the visualization achieved with constant fuzziness.

A  restrictive remark has to be made. We will not try
to answer the question posed initially in full generality. Instead of the
unknown driving potential $V(t)$ we will consider the special case of the atom
being subject to a $\pi-$ pulse of a resonant light field. Being
initially in the ground state $|E_1\ra$, the atom --- in the absence of
any measurement --- carries out a
transition to the upper state $|E_2\ra$ (Rabi transition). This is
the undisturbed motion to be visualized by means of our measurement scheme. 
The evaluation of
continuous fuzzy measurements is done by Monte
Carlo simulations. From the results obtained here we get insights for
the more general case of an unknown $V(t)$.  

The paper is organized as follows: Section 2 contains  a brief
description of the phenomenological model used to describe continuous
fuzzy measurements. In section 3 we define quantities that represent the
modification of the dynamics due to the influence of the measurement
on one hand and the reliability of the readouts on the other. These quantities are then
evaluated for measurements with constant fuzziness. In section 4
we investigate continuous measurements with time dependent fuzziness.
In section 5 continuous measurements with fuzziness depending on the
readout are studied. Both are compared with the measurements with constant
fuzziness.  The scheme of measurement introduced in section 5 can also be
used in the general case of unknown dynamics. We conclude with a
summary in section 6.

\section{Continuous fuzzy measurements}
We consider a two level atom submitted to a $\pi$-pulse of intensive,
resonant laser light. The respective Hamiltonian reads  
\begin{equation}
H =  E_1|E_1\ra\la E_1| + E_2|E_2\ra\la E_2| +v_0\left( \exp\{-i\omega
(t\} |E_2\ra\la E_1| + h.c. \right)\,,
\end{equation}
with $ \omega = (E_2 -E_1)/\hbar$ and $v_0 = \la E_2|{\mathbf
d}{\mathbf E}| E_1\ra$, where $ {\mathbf d }$ is the dipole moment of the atom
and $ {\mathbf E}  $ is the amplitude of the electric field
strength. The pulse lasts from $t=T_1$ untill $t=T_2$. If
no further influence is present, an atom in the ground state at $t=T_1$
performs a Rabi transition to the upper state  $|E_2\ra$ in the course of 
the $\pi$-pulse ($T_2-T_1 = \hbar\pi/v_0$).  Before and after the pulse the 
dynamics of the atom is governed by its free Hamiltonian
\begin{equation}
H_0=
E_1|E_1\ra\la E_1| + E_2|E_2\ra\la E_2|\,.
\end{equation}
A continuous
fuzzy measurement of energy (observable $H_0$) during the time interval $[0,T]$ containing $[T_1,
T_2]$ produces a readout $[E]$. 

Since we want to discuss a measurement applied to a single atom we
employ in the following a selective description of the measurement.
Given 
the initial state of the atom and given a
particular readout $[E]$, a theory describing continuous fuzzy  measurements has to
answer the following two questions: i) how has the state evolved during
the measurement and ii) what is the probability density $p[E]$ to
measure the readout $[E]$. The answers can be given 
in a phenomenological theory of continuous measurements. For a survey
of this theory see \cite{Men98revEng}. As already mentioned in the
introduction, these measurements find their realization by means
of sequences of unsharp measurements \cite{AudM98}.

Given a certain readout $[E]$, the  effective Schr\"odinger equation 
\begin{equation}
i\hbar\frac{\partial}{\partial  t} |\psi^{[E]}_t\ra = H_{[E]} |\psi^{[E]}_t\ra
\label{effecteqn}
\end{equation}
with complex Hamiltonian $H_{[E]}$
\begin{equation}\label{effect-Ham}
H_{[E]}\ = H - i\kappa\hbar \,\big( H_0 - E(t)
\big)^2
\label{effham}
\end{equation}
determines the unnormalized  
 solution $|\psi^{[E]}_t\ra$, thus  giving the answer to question i).
From this the probability density in question ii) can be
evaluated by  
\begin{equation}
p\left[ E\right]=\langle\psi_T^{[E]}|\psi_T^{[E]}\rangle \,
\label{prob}\
\end{equation}

The second term in the Hamiltonian (\ref{effham}) leads to  damping
of the amplitude of $|\psi_t^{[E]}\ra$. The amount of damping for
fixed  $\kappa$ depends on how close the readout $[E]$ is to the
curve of the expectation value  $\la H_0\ra$, for details see \cite{AudM97}. Large damping
implies because of (\ref{prob}), that the readout is improbable.   
In (\ref{effham}) $\kappa$ represents the strength of the
measurement. If $\kappa$ is small, such that $H$ dominates
$H_{[E]}$, the measurement perturbs the evolution only a little. If
$\kappa$ is great, the evolution is overwhelmed by the influence of
the measurement. 

The influence of the fuzzy measurement on the atom competes with the 
influence of the  external field driving the atom to the upper
level. The strength of the latter is characterized by half the Rabi period
$T_R/2$ - the time a Rabi transition takes without measurement.
In order to compare both influences, we  introduce a characteristic
time for the continuous measurement \cite{AudM97}.
 The effective resolution time  
\begin{equation}
  \T := \frac{4\pi}{\kappa\Delta E^2 } \,, 
\end{equation} 
may serve as a measure of fuzziness of a
continuous measurement. In what follows we
refer to  this quantity simply as {\it fuzziness} $\T$.

\section{Efficiency of visualization}
As already pointed out in the introduction, the visualization of a
single Rabi transition by a continuous measurement becomes efficient
if  two properties are fulfilled :
weak influence (later on specified as softness) of the measurement and
reliability of the  measurement readouts.
In this section we first introduce quantities to specify these
properties and then apply them to characterize measurements with
constant fuzziness. 
\subsection{Softness  and Reliability}
\label{concepts}
For a visualization the measurement  should be likely to only modify and not prevent
the Rabi transition. In order to fix what can be regarded as a
modified or approximate Rabi transition, we use the following,
somewhat minimal condition:
\begin{equation}
|c_2(T)|^2  > 0.5 
\end{equation}
for the component 
\begin{equation}
c_2(T)=  \frac{|\la E_2|
\psi_T^{[E]}\ra|^2}{\sqrt{\la\psi_T^{[E]}|\psi_T^{[E]}\ra}}\ge 0.5\,.
\end{equation}
The probability that the state will perform such an
approximate Rabi transition is given by
\begin{equation}
s := \int_{\cal U} p[E]d[E] \;\;\;\mbox{with}\; {\cal U}=\{[E]\mid
|c_2(T)|^2 \ge 0.5\}\;.
\end{equation} 
We refer to $s$ as {\it softness} of the continuous measurement. 

What can we read off from a readout [E] and how reliable is it? In
order to
answer these questions we turn to the realization given in
\cite{AudM98} of the
phenomenological scheme we are using. There it has been shown: if many
single weak measurements of energy are performed on the same
normalized states
$|\psi(t_0)\ra$, then the statistical mean value $\tilde E(t_0)$ of all
individual measurement results $E$ obeys the relation 
\begin{equation}
\frac{\tilde E(t_0) -E_1}{\Delta E} = \frac{\bar E(t_0) -E_1}{\Delta
E} = |c_2(t_0)|^2 
\end{equation}
with $\bar E(t_0) = \la\psi(t_0)| H_0| \psi(t_0)\ra$.
This has led to the idea to take $(E(t)-E_1)/\Delta E$ 
of a complete
single readout $[E]$ obtained under the influence of the driving
potential 
as an estimate of $|c_2(t)|^2$. How reliable is this?

Turning again to the total curves we introduce a mean deviation $d$
according to 

\begin{equation}
\label{d}
d^2 = \int p[E]d[E]\frac{1}{T}\int dt
\left[\frac{E(t)-E_1}{\Delta E} -|c_2(t)|^2\right]^2
 \;. 
\end{equation}
We call $r=1/d$ the {\it reliability} of a continuous measurement. It
indicates how reliable it is, that $(E(t)-E_1)/\Delta E$ of a single
readout agrees with $|c_2(t)|^2$.

\subsection{Constant fuzziness}
\begin{figure}
\centerline{
\epsfig{figure= 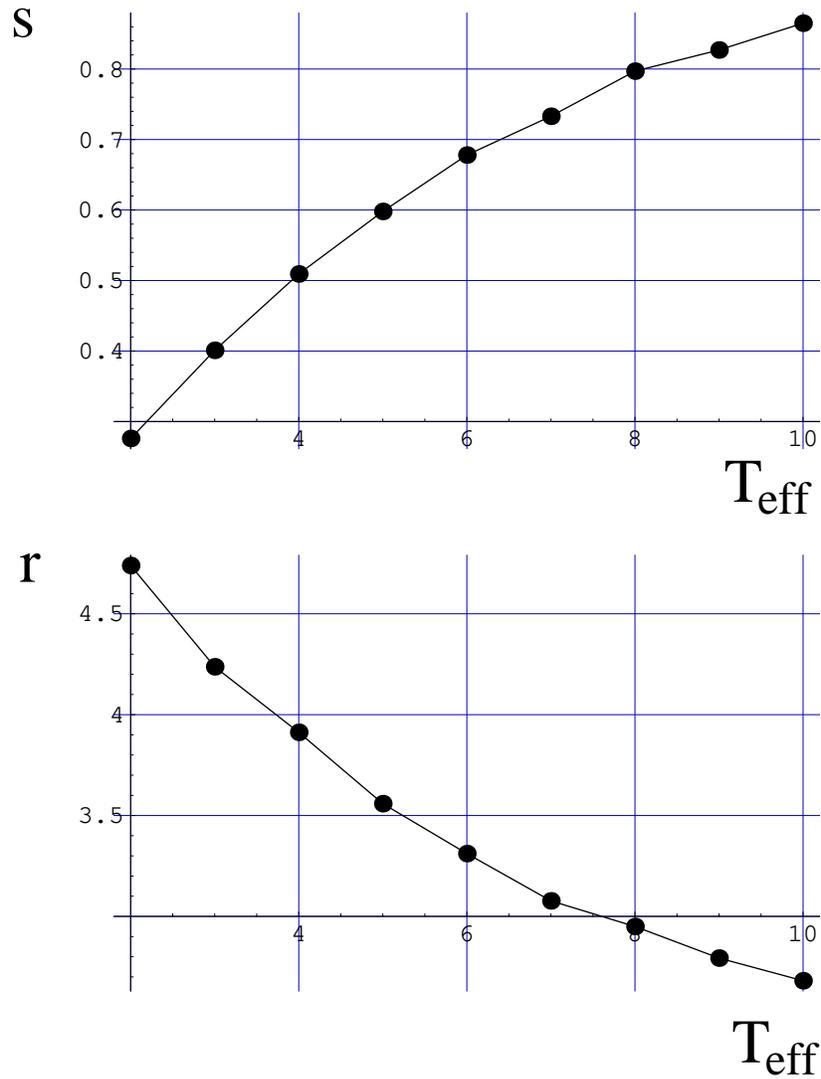, width=0.8\textwidth, angle=0}}
\caption{Upper part: softness $s$ of the influence of a continuous
measurement  for different values of constant fuzziness $\T$ (in units of
$T_R/3$). 
Lower part: The
correspondent reliability $r$ of a readout regarding the question  whether
it represents the evolution of $|c_2(t)|^2$} 
\label{pconst}
\end{figure}

We apply the concepts introduced in section \ref{concepts} to the
special case of measurements with fuzziness $\T$ kept constant during
the measurement (comp. \cite{AudM98,AudM97,AuMNam97}). The numerically obtained results
are displayed in Fig. \ref{pconst} and Fig. \ref{tcomp}. 
Fig. \ref{pconst} shows, how with growing fuzziness $\T$  softness $s$
increases and reliability $r$ decreases.
For low fuzziness we see the Zeno regime of strong measurement. The
 state motion differs largely from the Rabi transition, but is
well visualized. In the Rabi regime of high fuzziness the Rabi
transition of the state is preserved but cannot be visualized by the readout.
The relation between reliability and softness for measurements with
constant fuzziness can be seen from
Fig \ref{tcomp} (small circles). The pairs $(r,s)$ for different
values of fuzziness lie almost on a straight line.

Returning to our initial question we study a continuous measurement in the
intermediate regime.  For $\T=5$  an approximate transition takes place
in $60$ percent of the cases ($s=0.6$) and the mean deviation amounts to
$d=0.28$ in units of $\Delta E$. For a single measurement it is
therefore to be expected 
that the readout $[E]$ reflects approximately the evolution of
$|c_2|^2$. 

We now address the question whether an improvement of $r$ and $s$ can be
achieved if fuzziness varies in the course of the measurement.
      
\section{Time dependent fuzziness}

In order to improve the efficiency of the continuous fuzzy measurement
one may for example think of choosing a low fuzziness $\T$. This
increases the reliability $r$ but decreases softness $s$. If fuzziness $\T$ is
kept low over the whole measurement not much is gained because the
original Rabi transition is strongly modified.
But there are specific time intervals at which even a strong influence
of an energy measurement (small softness) is likely to modify a Rabi transition
only to a small amount. This is at the beginning ($t=T_1$) and at the end
($t=T_2$) of the $\pi$-pulse when the state of the undisturbed Rabi
transition  is  close to an energy eigenstate. Even a
projection measurement is not likely to modify the state very much 
in this case. Just the opposite situation can be found in the middle
of a $\pi$-pulse. Both observations lead to the idea to vary fuzziness
in time in order to take advantage of the varying sensitivity of the
system. In order to do so, one has to know the time
development of the state $|\psi(t)\ra$ beforehand, accordingly $V(t)$
must be known \footnote{An application for measurements with time
dependent fuzziness could be to visualize the motion of the state, in
case the potential $V(t)$ is known but it is not known if it  is turned on, i.e. $V=
V(t)$, or turned off
($V = 0$).}. We discuss  as above the real-time visualization of a
Rabi transition. 

\begin{figure}
\centerline{
\epsfig{figure= 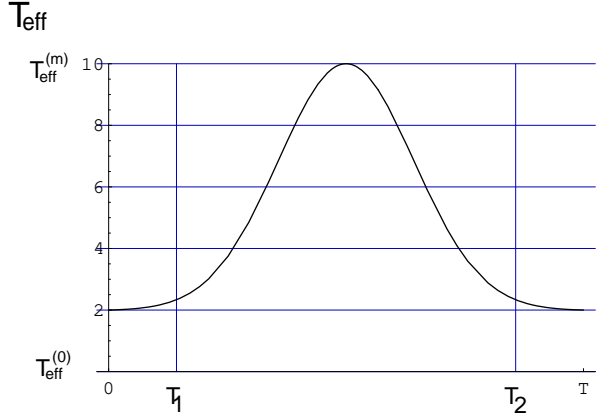, width=0.6
\textwidth, angle=0}}
\caption{Time dependent fuzziness $\T$ (in units of $T_R/3$) with width $\delta T=0.2$. \label{gaust}}
\end{figure}

We assume a time dependent fuzziness $\T(t)$ of Gaussian shape
(comp. Fig. \ref{gaust}) with width $\delta T$. The maximum $\Tm$ is
located at the middle of
the pulse at $t=(T_1+T_2)/2$. An offset value $\T^{(0)}\ne 0 $ is obtained at
$t=0$ and $t=T$. In order to see the influence  of a  varying width
$\delta T$, the maximum $\Tm$ and the minimum $\T^{(0)}$ are kept fixed
and the Gaussian  is rescaled
appropriately. This leads to

\begin{equation}
\T(t) =  a\lambda(t)+b
\end{equation}
with 
\begin{equation}
\lambda(t)  :=  \exp\left(\frac{(t-T/2)^2}{2(\delta T)^2}\right)\;
\end{equation}
and
\begin{equation}
a := \frac{\T^{(m)}-\T^{(0)}}{1-\lambda(T)}\;, \;\;b =:
\T^{(m)} - a \,.
\end{equation}

\begin{figure}
\centerline{
\epsfig{figure= 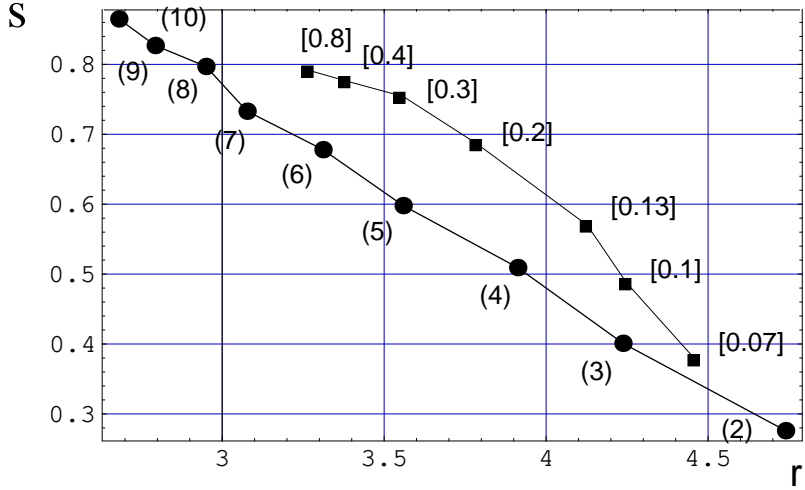, width=0.8\textwidth, angle=0}}
\caption{Comparison between constant  fuzziness ( small circles ,
the number in round brackets is to  value of $\T$) and  time
dependent fuzziness
(small squares, number  in  straight  brackets 
is the width $\delta T$ in
units of  $ T_{\rm R}/2$, $\Tm=10, \T^{(0)}=2$).
$s$ is the softness of a measurement and $r$ its reliability.
\label{tcomp}}

\end{figure}

We computed the reliability $r$ and softness $s$ of measurements with
fuzziness $\T(t)$  for different width $\delta T$. The results can be
seen in Fig. \ref{tcomp} where pairs of
$(r, s)$ are plotted for time dependent fuzziness (small
squares) and constant fuzziness (small
circles). We consider a measurement characterized by
$(r^{(1)},s^{(1)})$ to be 
``better than'' a measurement with  $(r^{(2)},s^{(2)})$ if both ---
the reliability and the softness of the first measurement ---
are greater: $s^{(1)} > s^{(2)}$ and $r^{(1)}>
r^{(2)}$.
%\footnote{Other definitions of better would involve a
%statement about whether a greater reliability or a greater softness
%is more valuable.}. 
In fact not all measurements are comparable in
terms of this definition
%\footnote{the definition of ``better than''
%induces a partial order.} 
but the ones
which are better than a particular one lie in the diagram on the right
and higher. The ones worse than the particular one lie to the left and
lower. The best measurement of the diagram would lie in its upper
right corner.

Fig. \ref{tcomp} shows that the results of  
measurements with constant fuzziness
can clearly be improved by using measurements with time dependent
fuzziness.
For example the measurement with time dependent fuzziness and width $\delta T = 0.2$ is
better than measurements with constant fuzziness with $4.3\le\T\le 6$. 
The former
measurement leads to $s = 69\%$ and $d = 0.26\Delta
E$. 
\\

Another way to display the results of the numerical calculations 
are density plots of the readouts $[E]$ and the curve of the 
squared component $|c_2(t)|^2$.  
The density plot of the readouts $[E]$ is obtained by dividing the
E-t-plane into squares of equal area 
and calculating  the probability that a smoothed curve [E]
(c.f. Appendix) crosses a particular
square.
\begin{equation}
P_S := \int_S P[E] d[E]\;, \mbox{with}\;\; S=\{ [E]| (t,E(t))\in
\mbox{square}\}\;.
\end{equation}

The degree of grayness of the square is then chosen according to $P_S$. The 
$|c_2(t)|^2$-density plot is created analogously, apart from the fact
that  the curves $|c_2(t)|^2$  do not have to be smoothed, since they do not
posses rapid oscillations. 

\begin{figure}
\centerline{
\epsfig{figure= 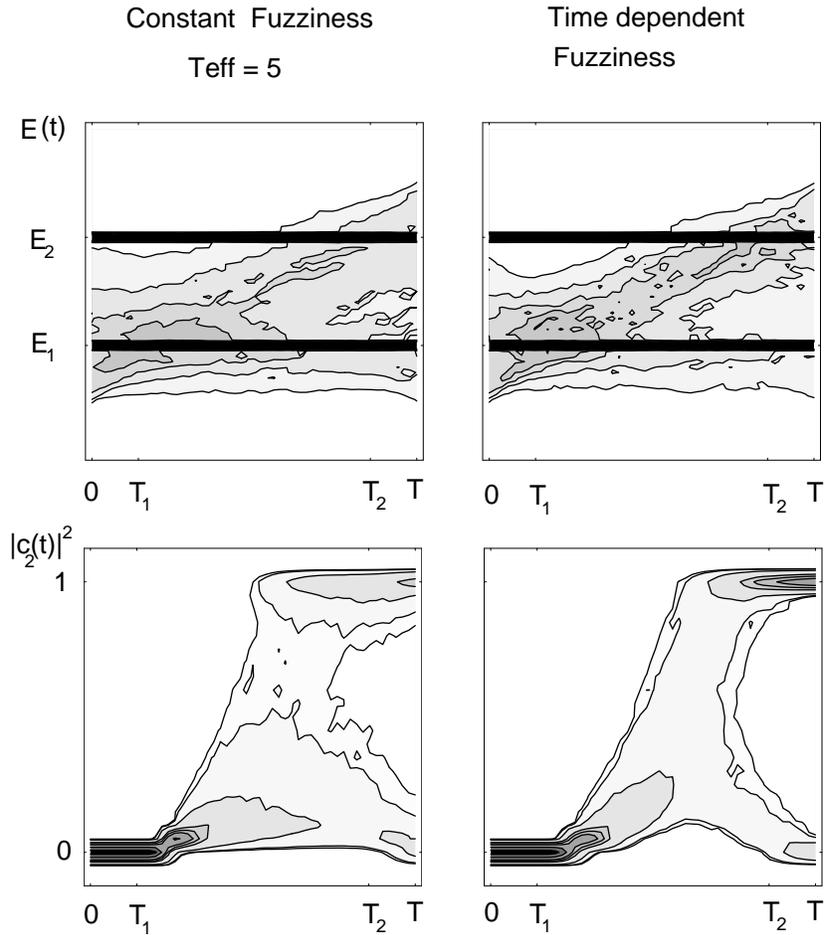, width=0.8\textwidth, angle=0}}
\caption{Visibility of the atomic transition: on the left with constant fuzziness ($\T=5$), on the right
 with time dependent  fuzziness $T_{\mathrm eff}(t)$ ($\delta T =
0.2$). The upper diagrams show density plots of 
measured energy curves (measured energy versus time). The lower
density plots display the corresponding curves of the squared
component $|c_2(t)|^2$ over time $t$.
Comparing constant with time dependent fuzziness, there is a progress
in terms of visibility of the transitions.}
\label{denst}
\end{figure}

In Fig. \ref{denst} density plots for constant fuzziness and for
time dependent fuzziness are displayed. 
The latter shows an improved correlation between  
readout $[E]$ and squared component $|c_2(t)|^2$. In particular we
can read off from the energy plots that
the area with the second smallest degree of grayness
(corresponding to $0.3\le P_S\le 0.4$) is more narrow for time
dependent fuzziness than for constant fuzziness. 
While in case of constant fuzziness 
there are still two branches of readouts (going up  and staying
down), for time dependent fuzziness there is only one strong branch
representing curves that show an approximate transition.  
The plots of the squared component $|c_2(t)|^2$ indicate a
greater softness for measurements with  time dependent fuzziness.
Both kinds of measurements show two branches, for time dependent
fuzziness the ends of both  branches are more narrow than for constant
fuzziness.

\section{Energy dependent fuzziness}

In the general case the driving potential $V(t)$ and therefore the
dynamics of the system without measurement are unknown. It will thus
not be possible to design beforehand an adjusted
time dependent fuzziness $\T(t)$. Nevertheless one can also in this
case take advantage of varying fuzziness.  Applying the same reasoning
as for time dependent fuzziness, we have to use a large fuzziness
(small perturbation by the measurement) when the system is not close to
an energy eigenstate, i.e. for $|c_2(t)|^2 \approx 1/2$. The value of the  readout
$E(t)$  is
correlated to $|c_2(t)|^2$ and $\la H_0\ra_t$. This means that 
energy dependent fuzziness $\T(E)$ should have  its maximum near
$(E_2-E_1)/2$ and should be low for $E_1$ and $E_2$.

\begin{figure}
\centerline{
\epsfig{figure= 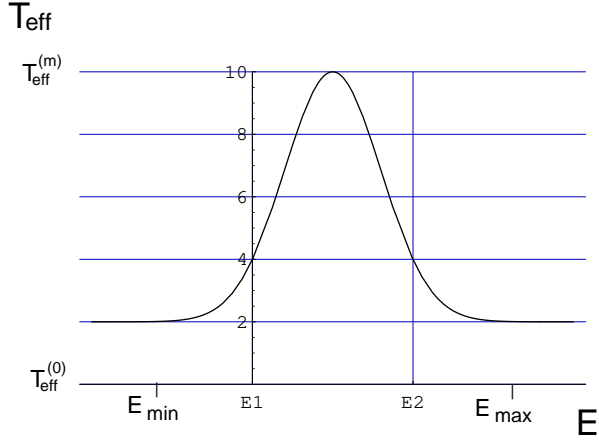, width=0.6
\textwidth, angle=0}}
\caption{Energy dependent fuzziness $\T$ (in units of $T_R/3$) with width $\delta E=0.3$}.
\label{gausen}
\end{figure}

We test this idea once more  for the case of a driving potential
which leads to a Rabi transition if no further influence is present.
For $\T(E)$ we assume a Gaussian   
$\T(E) =  a\lambda(E)+b$ with width $\delta E$ and 
\begin{eqnarray}
\lambda(E) & := & \exp\left(\frac{(E-(E_2+E_1)/2)^2}{2(\delta E)^2}\right)\;,\\
a& =& \frac{\T^{(m)}-\T^{(0)}}{1-\lambda(E_{\mathrm max})}\;, \;\;b = \T^{(m)} - a \,.\nonumber
\end{eqnarray}
We choose again as maximum $\Tm =10$ and as minimum $\T^{(0)}=2$, comp. 
Fig. \ref{gausen}.

\begin{figure}
\centerline{
\epsfig{figure= 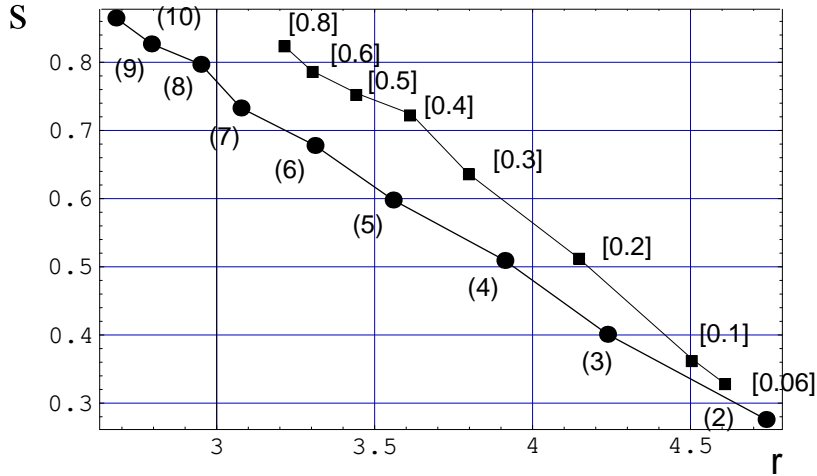, width=0.8\textwidth, angle=0}}
\caption{Comparison between constant fuzziness (small circles, the
number in round brackets is the value of $\T$)
and energy dependent fuzziness (small squares, the number  in straight
brackets is the width $\delta E$ in units of $\Delta E$, $\Tm =10,
\T^{(0)} =2$). $s$ is the softness of the measurement and $r$ its reliability.}
\label{ecomp}
\end{figure}

The results in terms of reliability and softness for different values of $\delta E$ (small squares) are
plotted in Fig. \ref{ecomp}. They can be compared with the values of $r$
and $s$ from constant fuzziness. 
As in the case of time dependent fuzziness, there is no improvement of 
measurements with  high or low constant fuzziness. But in the
intermediate regime, which is the important one for visualization, the results for
energy dependent fuzziness are better than for constant 
fuzziness. It is satisfying that in this regime not only with time dependent but also with
energy dependent fuzziness the results from constant fuzziness can be
improved. Time dependent fuzziness leads in the
intermediate regime to slightly higher values of reliability and
softness than energy dependent measurements.

\section{Conclusion}

The normalized state $|\psi(t)\ra
= c_1(t)|E_1\ra + c_2(t)|E_2\ra$ of a $2 $-level atom performs
Rabi oscillations under the influence of an external
driving field. It is assumed that there is only one realization of
this process. We ask the questions if it is possible  to obtain a
visualization of $|c_2(t)|^2$ in real-time using an appropriate
measurement scheme. It is shown that this is possible up to a certain
degree, if a continuous fuzzy measurement of energy is performed
leading to a readout $[E]$. In order to quantify  the efficiency of visualization for a given
fuzziness we have introduced for a measurement the complementary
concepts of softness (small disturbing influence) and reliability
(the curve of $|c_2(t)|^2$ agrees essentially with the readout $[E]$). It is discussed in detail how both
demands can be improved at the same time,  if fuzziness is chosen to
be appropriately time dependent or energy dependent.
\section{Acknowledgment}
This work has been supported in part by the Deutsche
Forschungsgemeinschaft and the Optik Zentrum Konstanz. 
\section*{Appendix: Numerical simulations}

\subsection*{Numerical simulations}
\label{numerical}
Since the effective Schr\"odinger equation (\ref{effecteqn}) does not
posses a closed form 
solution for general readout $[E]$, we simulated continuous fuzzy
measurements numerically.
The simulation can be divided into the following steps.

\begin{enumerate}
\item Generate a random curve $[E]$.
\item Insert $[E]$ into equation (\ref{effecteqn}) and solve them with
initial condition $|\psi_0\ra=|E_1\ra$. 
\item Compute $p[E]$.
\item Repeat steps 1.-3. $n$ times and process data.
\end{enumerate}

\noindent 
{\bf Description of the steps}\\
\noindent
Step 1.: Energy readouts are generated  out of the class of
functions
\begin{equation}
E(t) = g(t)+ \sum_{k=1}^{m}\,a_k\sin\left(\frac{k\pi t}{T}\right)\,,
\end{equation}
where  $g(t)$   is a straight line.
The initial and final value of the readout $E(0)=g(0)$ and $E(T)=g(T)$ are
chosen by random out of the interval $I_E:=[E_1-\Delta E/2,E_2+\Delta
E/2]$. In addition the Fourier coefficients $a_k$ are randomly taken
out of $I_a:=[-0.35\Delta E,0.35\Delta E]$. In the simulations we used
$m=10$, our results are stabile if higher Fourier terms are taken into
account ($m>10$).
\\
Step 2.: The effective Schr\"odinger equation  is
solved using computer algebra. The whole simulation was implemented in 
Mathematica [Wolfram Research]. In the computation of $|\psi_t^{[E]}\ra$
 we approximated $v(t)$ which describes the processes of turning on and
turning off the laser by a product of two smoothed step functions.\\ 
\\
Step 4.: Mean values of quantities $f$ that characterize a continuous
measurement with a certain fuzziness are calculated from the data of
the $n$ repetitions. They serve as approximation of the expectation value
of f:
\begin{equation}
\bar f = \int f[E] p[E]d[E]\approx   
\frac{1}{N}\sum_{i=1}^n\;p[E_i]f[E_i] \quad \mbox{with}\,\, N= \sum_{i=1}^n\;p[E_i]\,.\end{equation}

We chose  $n= 10^4$ in order to obtain  a
relative error of 
$\delta \bar f/\bar f \approx 1\% $. $n$ has to be increased for larger intervals $I_E,I_a$.\\
Density plots of the curves $|c_2(t)|^2$ and smoothed $[E]$ are made.  
$[E]$ is smoothed in order to extract information about the dynamics on
the scale of the order of the Rabi period $T_R$. The smoothing is done
by multiplying the Fourier coefficients $a_k$ with $exp\{-(k/3)^2\}$.
Thereby fast oscillations are damped. A real measuring apparatus
may not be able to display fast oscillations because of its inherent inertia.


\begin{thebibliography}{99}
%%% Bylines
\bibitem{Milburn}
M.J.Gagen and G.J.Milburn,
Phys. Rev. A47, 1467, 1993.

\bibitem{Zeno} B.Misra and E.C.G.Sudarshan,
J. Math. Phys. {\bf 18}, 756 (1977);
C.B.Chiu, E.C.G. Sudarshan and B.Misra,
Phys. Rev. {\bf D~16}, 520 (1977);
A.Peres, Amer. J. Phys. {\bf 48}, 931 (1980);
F.Ya.Khalili, Vestnik Mosk. Universiteta,
ser. 3, no.5, p.13 (1988);
W.M.Itano, D.J.Heinzen, J.J.Bollinger and
D.J.Wineland, Phys. Rev. {\bf A~41}, 2295 (1990);
A.Beige and G.C.Hegerfeldt,
Phys. Rev. {\bf A~53}, 53 (1996).

\bibitem{Ali74} S. Ali and G. Emch, J. Math. Phys. {\bf 15}, 176
(1974); P. Bush, M. Grabowski, P. J. Lahti {\em Operational Quantum
Physics}, Springer Verlag Heidelberg, 1995. 

\bibitem{ContinMeas}
H.D.Zeh,
Found. Phys. {\bf }1, 69 (1970); {\bf 3}, 109 (1973);
E. B. Davies, {\em Quantum Theory of
Open Systems}, Academic Press: London, New York, San Francisco,
1976;
M. D. Srinivas, {\em J.Math. Phys.} {\bf 18}, 2138 (1977);
A. Peres, Continuous monitoring of quantum
systems, in {\em Information Complexity and Control in Quantum
Physics}, ed. by A. Blaquiere, S. Diner, and G. Lochak, Springer,
Wien, 1987, pp. 235;
D. F. Walls and G. J. Milburn, Phys. Rev.
{\bf A 31}, 2403 (1985);
E. Joos, and H. D. Zeh, Z.Phys. {\bf B 59}, 223 (1985);
L. Diosi, Phys. Lett. {\bf A 129}, 419 (1988);
H.Carmichael,
{\em An Open Systems Approach to Quantum Optics},
Springer, Berlin and Heidelberg, 1993;
A. Konetchnyi, M. B. Mensky and V. Namiot,
Phys. Lett. {\bf A 177}, 283 (1993);
P.Goetsch and R .Graham, Phys. Rev. {\bf A~50}, 5242 (1994);
T.Steimle and G.Alber, Phys. Rev. {\bf A~53}, 1982 (1996);
M. B. Mensky, Phys. Rev. {\bf D 20}, 384 (1979);
Sov. Phys.-JETP {\bf 50}, 667 (1979); 
M. B. Mensky, {\em Continuous Quantum
Measurements and Path Integrals}, IOP Publishing: Bristol and
Philadelphia, 1993;
F.Ya.Khalili,
Vestnik Moskovskogo Universiteta,
ser. 3, v.{\bf 29}, no.5, p.13 (1988), in Russian;
M. Brune, S. Haroche, V.Lefevre, J. M. Raimond, and N. Zagury, 
Phys. Rev. Lett. 65, 976 ((1990);
Phys. Rev. {\bf A~45}, 3260 (1992);
V.B.Braginsky and F.Ya.Khalili,
{\em Quantum Measurement},
ed. Kip S.Thorne,
Cambridge University Press,
Cambridge,  1992;
N.Gisin, P.L.Knight, I.C.Percival and R.C.Thompson,
J. Modern Optics {\bf 40}, 1663 (1993);
K.Jacobs and P.L.Knight,
Phys. Rev.~{\bf A~57}, 2301 (1998);
A. Peres, {\em Quantum Theory: Concepts and
Methods}, Kluwer Academic Publishers, Dordrecht, Boston \&
London, 1993; R. Onofrio, C. Presilla, and U. Tambini, Phys.
Lett. {\bf A 183}, 135 (1993);
U. Tambini, C. Presilla, R. Onofrio, Phys.
Rev. {\bf A 51}, 967 (1995).

\bibitem{AudM98}
J.Audretsch, M.Mensky
``Realization scheme for continuous fuzzy measurement of energy and
the monitoring of a quantum transition''
Preprint: quant-ph/9808062

\bibitem{AudM97}
J.Audretsch and M.B.Mensky, Phys. Rev. {\bf A~56},  44 (1997).

\bibitem{AuMNam97}
J.Audretsch, M.Mensky and V.Namiot,
Phys. Lett. A237, 1 (1997).

\bibitem{Men98revEng}
M. B. Mensky, Physics-Uspekhi,
{\bf 41}, 923, (1998).       

\end{thebibliography}
\end{document}